\documentclass[sigconf,natbib=true]{acmart} 
\usepackage{graphicx}
\usepackage{amsmath}
\usepackage{tabularx}   % remove
\usepackage{booktabs}   % remove
\usepackage{algorithm}
\usepackage[noend]{algpseudocode}  % remove
\usepackage{hyperref}
\usepackage{xcolor}   % remove
\usepackage{listings}
\usepackage{adjustbox}  % remove
\usepackage{multirow}
\usepackage{multicol}
\usepackage{enumitem}

\usepackage[show]{chato-notes}

\newcommand{\dataset}{MIRA~}
\newcommand{\pub}{\texttt{Publications}}
\newcommand{\rd}{\texttt{Research Data}}
\newcommand{\var}{\texttt{Variables}}
\newcommand{\ins}{\texttt{Instruments \& Tools}}
\newcommand{\lib}{\texttt{GESIS Library}}
\newcommand{\web}{\texttt{GESIS Webpages}}

\copyrightyear{2026}
\acmYear{2026}
\setcopyright{cc}
\setcctype{by-nc-nd}
\acmConference[SIGIR '26]{Proceedings of the 49th International ACM SIGIR Conference on Research and Development in Information Retrieval}{July 20--24, 2026}{Melbourne, VIC, Australia}
\acmBooktitle{Proceedings of the 49th International ACM SIGIR Conference on Research and Development in Information Retrieval (SIGIR '26), July 20--24, 2026, Melbourne, VIC, Australia}
\acmDOI{10.1145/3805712.3808614}
\acmISBN{979-8-4007-2599-9/2026/07}

\begin{document}

\title{MIRA: An LLM-Assisted Benchmark for Multi-Category Integrated Retrieval}
% \titlerunning{MIRA: An LLM-Assisted Benchmark for Multi-Category IR}

\author{Mehmet Deniz Türkmen}
\affiliation{%
  \institution{GESIS -- Leibniz Institute for the Social Sciences, Cologne, Germany}
  \country{}
}
\email{deniz.tuerkmen@gesis.org}

\author{Suchana Datta}
\affiliation{%
  \institution{University College Dublin, Ireland}
  \country{}
}
\email{suchana.datta@ucd.ie}

\author{Dwaipayan Roy}
\affiliation{%
  \institution{Indian Institute of Science Education and Research, Kolkata, India}
  \country{}
}
\email{dwaipayan.roy@iiserkol.ac.in}

\author{Daniel Hienert}
\affiliation{%
  \institution{GESIS -- Leibniz Institute for the Social Sciences, Cologne, Germany}
  \country{}
}
\email{daniel.hienert@gesis.org}

\author{Philipp Mayr}
\affiliation{%
  \institution{GESIS -- Leibniz Institute for the Social Sciences, Cologne, Germany}
  \country{}
}
\email{philipp.mayr@gesis.org}

\author{Derek Greene}
\affiliation{%
  \institution{University College Dublin, Ireland}
  \country{}
}
\email{derek.greene@ucd.ie}

\renewcommand{\shortauthors}{Mehmet Deniz Türkmen et al.} % added, following the email from Sheridon received on 29th by Doi

\begin{abstract}

Users increasingly expect modern search systems to offer a unified interface that seamlessly retrieves information from diverse data sources and formats.
However, current information retrieval (IR) evaluation benchmarks have not kept pace with this development, primarily due to the lack of test collections that represent the diversity of contemporary search domains. 
We address this critical gap with MIRA, a novel benchmark based on
% , constructed from the user logs of 
a large-scale social science search platform. 
\dataset is designed for category-aware ranking across heterogeneous
% scholarly resource types encompassing four distinct scholarly 
categories -- \pub, \rd, \var, and \ins~-- within a single, unified evaluation framework.
The proposed collection is distinctive in several ways: (1) it is built upon real user queries, providing a more realistic basis for evaluation; (2) it covers scholarly items from four distinct categories, enabling multi-faceted evaluation; and (3) it leverages a Large Language Model to generate topic descriptions
and narratives, 
as well as for relevance assessment with respect to these topics, substantially reducing the labor and cost of test collection generation.
% Creating such a collection through traditional manual annotation is prohibitively expensive, requiring expert assessors for each category. 
% Our further contribution is leveraging a Large Language Model (LLM) to overcome this bottleneck, using it to automatically generate comprehensive topic descriptions and, most critically, to provide scalable relevance judgments across all categories.
We release this resource
% All related resources can be found in a GitHub repository\footnote{\label{github_link}\footnotesize \url{https://github.com/suchanadatta/MIRA-LLM-Assisted-Benchmark.git}}.
%\footnote{\footnotesize \url{https://github.com/suchanadatta/MIRA-LLM-Assisted-Benchmark.git}} 
to benefit the community by providing a foundational testbed for the research on multi-faceted, category-aware, integrated, or cross-category information retrieval.\footnote{\label{github_link}\footnotesize \url{https://github.com/suchanadatta/MIRA-LLM-Assisted-Benchmark.git}}


\end{abstract}




\maketitle              % typeset the header of the contribution

\setcounter{footnote}{0}

\section{Introduction} \label{sec:intro}

In recent years, the paradigm of information retrieval (IR) has shifted steadily from siloed, single-type search experiences (e.g., ``find a document'') towards more integrated, multi-faceted user journeys in which users expect to pose a query and retrieve across heterogeneous content types~\cite{epasto-2014-reduce}.
Current IR evaluation benchmarks often do not reflect this complexity, primarily due to a shortage of test collections that encompass such a multi-categorical environment \cite{chenair}.
This limitation hinders the comprehensive assessment of IR systems in real-world scenarios~\cite{thakur2021beir}.
It is particularly evident in digital libraries and scholarly search platforms, where a single information need may be satisfied by a combination of research papers, underlying datasets, measurement variables, or specialized software tools~\cite{kacprzak-2017-querylog}.
Users entering a query might reasonably expect to see relevant research artifacts -- all in the same search interface.
However, existing IR evaluation benchmarks have remained largely monolithic, targeting a single item type -- such as passages, documents, or answers -- and rarely mirroring the cross‐category integration characteristic of modern search environments.
The shift towards a multi-faceted information landscape requires IR systems capable of effectively navigating and ranking items across multiple categories simultaneously.

The advancement of IR systems is intrinsically linked to the availability of high-quality, realistic evaluation benchmarks~\cite{sanderson2010test}. 
Historically, the IR community relied on Cranfield-style test collections -- comprising documents, topics, and relevance judgments -- to drive progress in a controlled and comparable manner~\cite{cleverdon}. 
While invaluable, many established benchmarks, such as the TREC ad-hoc tracks~\cite{voorhees2005trec} and CLEF campaigns~\cite{peters2010clef}, have often been constrained to a single predominant document type, such as newswire articles or web pages. 
This creates a critical gap between the simplified environments used for evaluation and the complex, multi-categorical reality of contemporary search platforms. 
Without test collections that mirror this integrated landscape, it is impossible to robustly evaluate and compare the performance of next-generation IR systems designed for it.

The challenge of creating such multi-categorical benchmarks is non-trivial.
The process is traditionally labor-intensive and expensive, requiring significant human effort to create search topics and, most critically, to annotate relevance judgments~\cite{zobel1998reliable}.
This cost is multiplied when a collection spans disparate item types, as the relevance criteria for a research dataset, for example, differ substantially from those for a scholarly publication~\cite{koesten2017trials}.
Consequently, the community lacks resources to study cross-category retrieval, result fusion, and the nuanced user behavior associated with complex search tasks spanning multiple information types.

Recent advancements in Large Language Models (LLMs) offer a promising opportunity to address the scalability challenges of test collection creation. LLMs have demonstrated strong capabilities in text understanding and generation, and their application to IR tasks has become an active area of research.
Prior work has explored using LLMs for tasks, such as query expansion~\cite{xia-etal-2025-knowledge} and synthetic document generation~\cite{rahmani2025syndl}.
More relevantly, initial investigations have begun to assess the potential of LLMs as relevance judges, with studies comparing their annotations to those of human assessors~\cite{faggioli2023perspective}. While these studies caution against directly replacing human judgment in all scenarios, they do highlight the potential of LLMs as a powerful means to augment and scale the curation process, particularly for constructing novel benchmarks where traditional methods are prohibitively expensive.

In this paper, we address a critical gap in IR evaluation by introducing a novel, multi-categorical test collection built from the user logs of a large-scale social science search platform.
Our benchmark is designed for the new paradigm of category-aware integrated search, encompassing four distinct scholarly categories — \pub, \rd, \var, and \ins~— within a single, unified evaluation framework. The proposed test collection is distinctive in several ways as follows:
\begin{enumerate}[leftmargin=*]

\item It provides comprehensive coverage of four distinct scholarly item categories, enabling a multi-faceted evaluation of integrated IR systems.

\item It is grounded in real user queries, ensuring that the resulting evaluation topics reflect authentic information needs.

\item It innovatively leverages an LLM to streamline the generation of topic descriptions and narratives, and to annotate item relevance. This approach significantly enhances the efficiency of the test collection generation process, reducing labor and costs, and offers a viable model for future benchmark development.
\end{enumerate}

\section{Related Work} \label{sec:rel-work}

Our work lies at the intersection of several key areas in IR research: the evolution of test collections, the specific challenges of multi-categorical search, and the emerging use of LLMs to automate and scale evaluation processes. This section reviews the relevant literature in these domains.

% \subsection{The Cranfield Paradigm and Evolving Test Collections}

\paragraph{\textbf{Cranfield Paradigm and Evolving Test Collections}.} 
The foundation of  modern IR evaluation is the Cranfield paradigm, which
% establishes the methodology of using 
introduced the use of test collections -- a \emph{document corpus}, \emph{topics}, and \emph{relevance judgments} -- to compare the 
% performance of different 
retrieval systems in a controlled manner~\cite{cleverdon}.
% This model has been instrumental in the field's progress, exemplified by large-scale campaigns such as 
This model has driven major progress through large-scale campaigns such as the TREC \cite{trec},
% \footnote{\url{https://trec.nist.gov/}}, 
CLEF \cite{clef},
% \footnote{\url{https://www.clef-initiative.eu/}}, 
NTCIR \cite{ntcir}
% \footnote{\url{https://research.nii.ac.jp/ntcir/index-en.html}}, 
, and FIRE \cite{fire},
% \footnote{\url{https://fire.irsi.org.in/}}.
% These initiatives have 
producing invaluable resources for evaluating ad-hoc retrieval~\cite{voorhees2005trec}, web search~\cite{trecdl}, complex question-answering~\cite{olvera2015qa} and many other tasks.
However, as search applications diversify, the limits of single-category collections have become evident.
Benchmarks like TREC-CAR~\cite{dietz2017car} 
% for complex answer retrieval or the 
and MS MARCO~\cite{craswell2021msmarco} mark progress but still focus on retrieving a single information type (e.g., text passages).
% for machine reading comprehension represent significant advances, they largely remain within the paradigm of retrieving a single type of information unit (e.g., text passages).
% The community has long recognized that the creation of such test collections is a resource-intensive process, requiring significant human labor for topic creation and relevance assessment, which acts as a bottleneck for the development of new evaluation
Creating such collections remains resource-intensive, requiring extensive human effort for topic design and relevance assessment, which constrains the development of new resources~\cite {zobel1998reliable,sanderson2010test}. 
% Recently, there has been discussion about how the static test collection approach can be improved to accommodate real-world search systems, which are dynamic and constantly changing
Recent work has explored ways to improve the static test-collection model to better reflect the dynamics of real-world search systems~\cite{keller2025}.

% \subsection{Test Collections for Multi-Categorical Search}

\paragraph{\textbf{Test Collections for Multi-Categorical Search.}}
Scholarly search demands multi-categorical retrieval, with information needs spanning publications, datasets, code, and other research objects~\cite{koesten2017trials}. 
Existing benchmarks address only limited aspects: ACL OCL includes papers, authors, and venues but is confined to computational linguistics~\cite{rohatgi2023acl}; TREC Chemical retrieves documents and patents but not heterogeneous item types~\cite{lupu2011trecchem}; and MMDocIR targets multimodal document elements rather than distinct object categories~\cite{dong2025mmdocir}. 
Dataset search has been examined separately through log analyses and tailored relevance criteria (such as~\cite{kacprzak-2017-querylog},~\cite{chapman2020datasetsearch}, and~\cite{kolyada2025dataset}), yet these efforts treat datasets in isolation. 
Our work builds on this understanding of domain-specific relevance and extends it by integrating multiple, distinct scholarly categories into a single benchmark.
To the best of our knowledge, no publicly available benchmark combines four heterogeneous research object types within a unified evaluation framework derived from real user logs.

\paragraph{\textbf{Leveraging Real User Logs.}}
% Constructing test collections from real user query logs is a established methodology to ensure ecological validity and mitigate the biases inherent in laboratory-created topics
% Constructing test collections from real user logs is an established way to enhance ecological validity and reduce biases introduced by lab-created topics~\cite{alonso2012crowdsourcing}.
Constructing test collections from real user logs enhances ecological validity by capturing authentic, under-specified information needs~\cite{alonso2012crowdsourcing}.
Foundational analyses, such as those of the AOL log~\cite{pass2006aollog}, revealed the complex and often under-specified nature of genuine information needs.
% Analyses, like the AOL log study~\cite{pass2006aollog}, revealed the complex, often under-specified nature of real information needs.
% This approach is particularly critical in specialized domains like scholarly and professional search, where log studies of digital libraries~\cite{jones2000,sun2008personalized,hienert2017user,fu2021} and data repositories~\cite{kacprzak-2017-querylog} have detailed the complex, iterative search behaviors of experts.
% This approach is vital in specialized domains such as scholarly and professional search, where log studies of digital libraries~\cite{jones2000,sun2008personalized,hienert2017user,fu2021} and data repositories~\cite{kacprzak-2017-querylog} have shown experts' iterative search behaviors.
This approach is vital in scholarly and professional search, where log studies of digital libraries~\cite{jones2000,sun2008personalized,hienert2017user,fu2021} and data repositories~\cite{kacprzak-2017-querylog} reveal iterative expert behaviors.
The main challenge, however, is to obtain reliable large-scale relevance judgments. A common solution is to use implicit feedback as a relevance proxy~\cite{jung2007,zhang2010}, exemplified by MS MARCO~\cite{craswell2021msmarco}, which paired Bing queries with clicked passages to produce weak labels at scale.
% However, the primary challenge in log-based collection building is obtaining reliable relevance judgments at scale. A dominant solution to this is the use of implicit feedback as a proxy for relevance~\cite{jung2007,zhang2010}.
% The MS MARCO dataset~\cite{craswell2021msmarco} is a landmark example, where real Bing queries were paired with clicked passages to create weak relevance labels, enabling the creation of massive-scale resources.
% This paradigm has been successfully adapted to specialized domains, such as the creation of test collections from clinical search logs for systematic reviews~\cite{scells2017collection} and the use of real legal questions in the COLIEE benchmark~\cite{kim2023coliee}, demonstrating the robustness of this approach for complex, professional search environments.
This paradigm extends to domains such as clinical search~\cite{scells2017collection} and legal question answering in COLIEE~\cite{kim2023coliee}, proving its effectiveness for complex, professional retrieval.
% 
% Our work is firmly situated within this tradition of log-driven, ecologically valid test collection construction.
% We derive our initial topic set from the logs of a large-scale social science search platform, ensuring that the core information needs are authentic and representative of its real user base.
% Our work follows this log-driven, ecologically valid tradition. We derive topics from logs of a large-scale social science search platform, ensuring that the information needs are authentic and representative.
% This grounds our multi-categorical benchmark in real researcher challenges, advancing evaluation beyond synthetic settings toward a more realistic basis for integrated retrieval assessment.
Following this tradition, we derive topics from logs of a large-scale social science search platform, grounding our multi-categorical benchmark in authentic researcher challenges and advancing evaluation beyond synthetic settings toward realistic, integrated retrieval assessment.
% This grounds our multi-categorical benchmark in the practical challenges faced by researchers, moving the evaluation beyond a purely synthetic or laboratory-based setting and providing a more realistic foundation for assessing integrated retrieval systems.

% \subsection{Large Language Models for IR Evaluation}

% The recent advent of powerful Large Language Models (LLMs) has opened new frontiers for automating and scaling IR tasks.
% Their application to evaluation is a rapidly growing subfield.

\paragraph{\textbf{Large Language Models for IR Evaluation.}} 
% LLMs have transformed the information retrieval landscape~\cite{zhu2025llmforirsurvey}, driving major advances from document ranking~\cite{qin2024large} to system evaluation~\cite{gao2025llm}. Their application to IR evaluation marks a paradigm shift, addressing the long-standing cost and scalability limits of the Cranfield paradigm~\cite{saracevic2008effects,zhu2023relevance,arabzadeh2025benchmarking}. A key research focus is using LLMs as relevance assessors to test whether they can mirror human judgment. Studies show mixed outcomes: Faggioli et al.~\cite{faggioli2023perspective} present a comprehensive framework highlighting that while LLMs often align well with human judgments on clear topics, their performance depends strongly on prompt design and task complexity.
LLMs have transformed IR, advancing document ranking, system evaluation, and addressing the long-standing scalability constraints of the Cranfield paradigm~\cite{saracevic2008effects,zhu2023relevance,arabzadeh2025benchmarking,zhu2025llmforirsurvey,qin2024large,gao2025llm}.
A key research direction examines LLMs as relevance assessors capable of mirroring human judgment. 
Findings are mixed: Faggioli et al.~\cite{faggioli2023perspective} demonstrate that while LLMs align well with humans on clear topics, performance is highly sensitive to prompt design and task complexity.
% LLMs have revolutionized the information retrieval landscape~\cite{zhu2025llmforirsurvey}, catalyzing significant advancements in key areas from the basic document ranking~\cite{qin2024large} to evaluation of systems~\cite{gao2025llm}.
% The application of LLMs to the core tasks of IR evaluation represents a paradigm shift aimed at overcoming the historic bottlenecks of cost and scalability associated with the Cranfield paradigm~\cite{saracevic2008effects,zhu2023relevance,arabzadeh2025benchmarking}.
%
% A primary research thrust has been the direct use of LLMs as relevance assessors, investigating whether they can replicate human judgment. Seminal studies in this area have yielded nuanced findings.
% Faggioli et al.~\cite{faggioli2023perspective} provide a comprehensive framework for understanding the potential and pitfalls, noting that while LLMs can achieve high agreement with human assessors on well-defined topics, their performance is sensitive to prompt design and task difficulty.

% Beyond direct relevance judgment, LLMs are being leveraged to automate other costly steps in the test collection lifecycle.
% This includes the generation of synthetic topics and narratives to expand existing collections or simulate new search scenarios~\cite{bonifacio2022inpars,wang2024sotopia}.

Beyond direct relevance assessment, LLMs are increasingly used to automate test collection development -- generating synthetic topics and narratives to expand collections or simulate new scenarios~\cite{bonifacio2022inpars,wang2024sotopia}. 
Their instruction-following capability also aids in creating assessment guidelines and judgment explanations, improving assessor training and consistency~\cite{liu2023argugpt,category-eval}. 
We combine these approaches by employing an LLM to generate topic descriptions, narratives, and initial relevance annotations across four scholarly categories, mitigating labor demands while leveraging scalability and established best practices.

Following~\cite{hienert2019digital}, we used dedicated user interactions (e.g., \texttt{views}, \texttt{downloads}) as initial relevance indicators. 
However, both LLM-based and human analyses revealed frequent mismatches with topical relevance (e.g., viewing irrelevant records).
We therefore adopted a hybrid approach in which the LLM assisted human assessors in producing final explicit judgments, rather than treating implicit signals as direct proxies.
% Following~\cite{hienert2019digital}, we used dedicated user interactions (e.g., document views, full-text or data downloads) as initial relevance indicators. 
% Yet, both LLM-based and human analyses revealed frequent mismatches with topical relevance (e.g., viewing irrelevant records). Therefore, we adopted a hybrid approach in which the LLM assisted human assessors in producing final explicit judgments, rather than treating implicit signals as direct proxies for relevance.

% \cite{arabzadeh2025benchmarking} - 

%%% Local Variables:
%%% mode: latex
%%% TeX-master: "main"
%%% End:

\section{GESIS Search: An Integrated Search System} \label{sec:gesis-search}

% 
% Our test collection is constructed from the user logs of GESIS Search~\cite{gesis-search}, an integrated digital library and search system tailored for the social sciences~\cite{hienert2019digital}.
% The platform is specifically designed to support the entire research lifecycle, from study design and data collection to analysis and publication, by providing a unified interface for a diverse set of scholarly resources.
% It enables social scientists to find, access, and interlink crucial resources such as research datasets, scientific publications, survey variables, and methodological instruments. The target audience for GESIS Search is social scientists and researchers from related disciplines. GESIS Search has approximately $30,000$ unique users per month with $200,000$ page impressions.
% 
Our test collection is derived from the user logs of GESIS Search~\cite{gesis-search}, an integrated digital library and search platform for the social sciences~\cite{hienert2019digital}. GESIS Search is designed to support the entire research life-cycle — from study design and data collection to publication — it provides a unified interface to diverse scholarly resources, enabling users to find and interlink datasets, publications, survey variables, and methodological instruments. The platform serves social scientists and related researchers, attracting about $30,000$ unique users with $250,000$ page impressions per month.

A key feature of GESIS Search is its rich, interlinked metadata model.
Information items are not isolated; they are connected through semantic relationships, allowing users to navigate the scholarly ecosystem.
For instance, a user can discover a publication and then directly navigate to the underlying research dataset it analyzes, or examine the specific survey variables and questionnaires used to collect the data.
This network of links facilitates exploratory search and enhances scientific transparency and reproducibility.
The platform supports this with advanced search filters, enabling users to refine their searches across different resource types, study collections, and recommendations to related content.
Furthermore, the ability to directly download datasets and materials empowers researchers to reuse existing data for their own research questions.

The content in GESIS Search is organized into six primary categories that represent the core information needs of empirical social science research.
The scale and specificity of these collections make it an ideal foundation for building a multi-categorical IR test collection. The categories are as follows:
\begin{itemize}[leftmargin=*]
    \item \textbf{\rd}: Approximately $7,700$ quantitative social science research datasets, with temporal coverage from $1945$ to $2026$ and a primary geographic focus on Germany and Europe.
    \item \textbf{\var}: A fine-grained collection of approximately $1.4$ million survey variables extracted from questionnaires. Roughly $200,000$ of these are high-quality variables, accompanied by detailed metadata including full question texts, answer categories, and frequency tables.
    \item \textbf{\ins}: Around $600$ methodological resources to aid in survey design and data collection. This includes pretested questionnaires, validated response scales, and analysis syntax files for statistical software, alongside guides for collecting survey and digital behavioral data.
    \item \textbf{\pub}: A corpus of approximately $260,000$ publications sourced from open-access repositories and literature that is explicitly linked to research data, supporting literature reviews and meta-analyses.
    \item \textbf{\lib}: A specialized collection of approximately $122,000$ publications from the GESIS internal library, with a strong focus on empirical social research and applied computer science.
    \item \textbf{\web}: Roughly $2,700$ informational pages from the GESIS website, detailing consulting services, training workshops, and other support offerings for social scientists.
\end{itemize}
For the purpose of constructing our multi-categorical benchmark, we focus on the four core scientific resource types: \textbf{\pub}, \textbf{\rd}, \textbf{\var}, and \textbf{\ins} that are also represented in the GESIS KG~\cite{SDN-10.7802-2969} with their semantic relationships. This selection captures the essential, interlinked dimensions of social science research.

\section{MIRA: A Multi-Category Integrated Retrieval Assessment Dataset} \label{sec:dataset}

To address the gap in evaluating integrated search systems, we introduce \textbf{MIRA}, a \textbf{M}ulti-Category \textbf{I}ntegrated \textbf{R}etrieval \textbf{A}ssessment test collection. 
The collection is constructed from real-world user interactions and data from the GESIS Search platform, providing a realistic and multi-faceted benchmark for scholarly IR. 
This section details its core components: the document collection, topics, and relevance judgments -- and outlines its availability.
The metadata is available under a Creative Commons Attribution 4.0 license, while the accompanying software is distributed under a GNU General Public License v3.0.
All resources are publicly available in the project’s GitHub repository, including prompt templates, generation parameters, pooling strategies, relevance judgments, and hyperparameter settings, along with all code and a quickstart guide for using the dataset.

\subsection{Collection}
The MIRA collection is a snapshot of the four core scholarly resource types within GESIS Search (see Section~\ref{sec:gesis-search}), encompassing a diverse set of social science information. The collection contains metadata for $7,634$ research datasets, $206,434$ high-quality metadata variables, $604$ instruments \& tools, and $254,097$ publications totaling $468,769$ documents, provided as a set of JSON files.

\begin{comment}
The collection statistics are summarized in Table~\ref{tab:col-stats}.

\begin{table}[h]
  \centering
  \caption{Statistics of the MIRA document collection}
  \label{tab:col-stats}
  \begin{tabular}{@{~}l@{~~~}|@{~}r@{~}}
    \hline
    \textbf{Category} & \textbf{\# Documents} \\
    \hline
    \rd & $7\,634$ \\
    \var & $206\,434$ \\
    \ins & 604 \\
    \pub & $254\,097$ \\
    \hline
    \textbf{Total} & \textbf{$468\,769$} \\
    \hline
  \end{tabular}
\end{table}
\end{comment}

% \begin{figure*}[!htb]
% \centering
% \includegraphics[width = 0.65\textwidth]{figures/top50_wordcloud.png}
% \caption{Word cloud of the top 50 topics derived from topic modelling.}
% \label{fig:top50_wordcloud}
% \end{figure*}

\begin{figure}[!h]
\centering
\includegraphics[width=0.8\columnwidth]{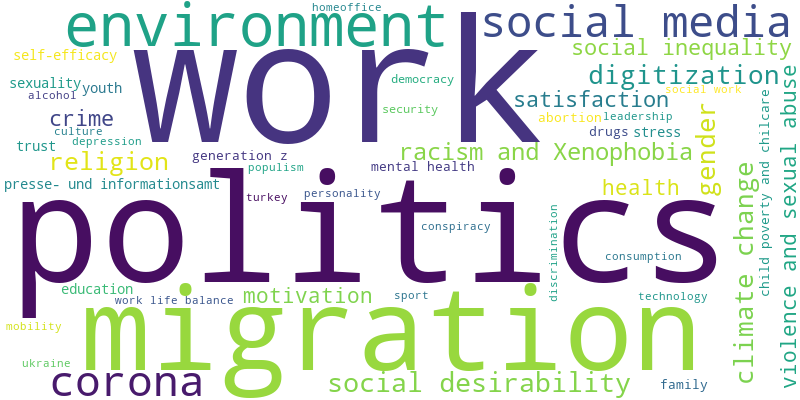}
\caption{\small
% Word cloud of top 50 topics derived from topic modeling.
Top-50 topic word cloud from topic modeling.
}
\label{fig:top50_wordcloud}
\end{figure}

\subsection{Topics}

The topics in \dataset originate from real user queries submitted to the GESIS Search platform. 
We used user logs collected between $2017$ and $2024$, comprising $16,335,937$ interactions that capture various user interactions during their search sessions. 
To identify interactions that indicate user interest and potential relevance between a query and a resource, we considered three action types (taking a cue from~\cite{hienert2019digital}): \emph{\texttt{View record}} (a click to show the detailed view of a record), \emph{\texttt{Download}} (downloading data or documents), and \emph{\texttt{Export}} (exporting citation information). 
Based on these actions, we extracted corresponding query–item pairs, resulting in $412,032$ pairs across the four categories in the data.%\rd, \var, \ins, and \pub~categories.

% A primary challenge in constructing a test collection from such logs is the inherent noise and redundancy of query data, where users express similar information needs with different terminology, languages, and specificity. 
The initial pool of multi-category queries contained significant semantic overlap, as users frequently expressed core information needs through multiple query variants.
Therefore, we performed query clustering to group these variations, ensuring that the final set of topics is both comprehensive and non-redundant.
To transform the raw query-item pairs into a coherent set of non-redundant evaluation topics, we performed clustering on the $412,032$ pre-selected queries using \texttt{BERTopic}~\cite{grootendorst2022bertopic}. 
Query embeddings were generated with the multilingual model, 
\textit{paraphrase-multilingual-MiniLM-L12-v2}~\cite{reimers2019sentence} to handle our mixed German–English dataset, producing 384-dimensional vectors. These vectors were further reduced to five dimensions using UMAP to improve clustering accuracy and efficiency, and were then grouped using the HDBSCAN~\cite{Campello2013HDBSCAN}.

This process effectively clustered semantically-related queries (e.g., `environment’ vs. `sustainability’), topical variants (e.g., `migration in Germany’ vs. `migration in the UK’), and translations (e.g., `Beruf’ vs. `job’) into coherent topics. After removing known-item searches (e.g., `ALLBUS’), we retained the 50 most frequent clusters for further analysis (see Figure~\ref{fig:top50_wordcloud}). 
% From this refined pool of candidate clusters, we selected the most frequent queries issued across all four target categories, ensuring that each topic in \dataset corresponds to an information need with demonstrated relevance across diverse resource types. In addition, we selected queries containing more than two words, as longer queries tend to express more specific information needs and therefore pose a more challenging retrieval task. By combining the most frequent queries executed across all four categories with longer, more specific queries, we aimed to construct a test collection that supports robust multi-category evaluation while also enabling clearer differentiation between retrieval effectiveness levels. 
From the refined clusters, we selected the most frequent queries across all four categories and restricted them to those with more than two terms to ensure specificity. This produced a test collection that supports robust multi-category evaluation and clearer differentiation in retrieval effectiveness.
As a result, we obtained a set of $200$ topics for our test collection, out of which $145$ are in German and rest $55$ are in English.
%From this refined pool of candidate topics, we selected only those where the original queries had been explicitly executed across all four target categories using the platform's (i.e. GESIS Search) category filter. This final selection criterion guarantees that each topic in \dataset represents an information need with demonstrated potential relevance to diverse resource types, making it suitable for robust multi-category evaluation, and thus we obtain a set of $215$ potential topics.
% 
For each resulting topic, we create a structured representation that includes the original query, a full description, and a detailed narrative. The description and narrative are uniquely created for each category in our dataset. These descriptions and narratives are generated using an LLM, guided by a purpose-built prompt designed to produce category-aware topic expansions. The exact prompt template and generation parameters are available in the project GitHub repository to ensure full reproducibility.
A sample snippet for the topic `immigration' is shown in Figure~\ref{fig:topic}.

\begin{figure}[h!]
    \centering
\begin{lstlisting}
<top>
<num>221</num>
<title>immigration</title>
<publication>
	<desc>Scholarly and policy publications on international immigration: migration flows, immigrant integration, asylum/refugee issues, labour migration, remittances,...</desc>
	<narr>Include empirical studies, reviews, policy analyses and theoretical works that address international migration, immigrant integration and assimilation, labor market outcomes of immigrants, refugee/asylum issues,...</narr>
</publication>
<research_data>
	<desc>Datasets relevant to immigration research: population registers, censuses and microdata with country-of-birth/citizenship, labour force surveys, asylum and...</desc>
	<narr>Include population registers, immigration and naturalization administrative data, household and labor force surveys with migrant identifiers (country of birth, citizenship, year of arrival),...</narr>
</research_data>
<variables>
	<desc>Variable metadata for migration concepts: country/region of birth, citizenship, legal status, reason for migration, year/age at arrival, duration, language proficiency,...</desc>
	<narr>Include variables such as migrant status (immigrant, emigrant, refugee, asylum-seeker), country of birth, citizenship, parental country of birth,...</narr>
</variables>
<instruments_tools>
	<desc>Survey modules, question batteries, scales and tools designed to measure migration status, integration, acculturation, legal status, and migration...</desc>
	<narr>Include survey modules and question batteries for migration (e.g., migrant background question sets used in EU surveys), language proficiency assessments, integration scales (social, economic integration indices),...</narr>
</instruments_tools>
</top>
\end{lstlisting}
      \caption{\small A topic (`immigration') description from the \dataset dataset.}
  \label{fig:topic}
\end{figure}

\subsection{Relevance Judgments}

A notable feature of the \dataset dataset is its scale and the methodology used for its relevance judgments.
For each of the $200$ selected topics, we identified a pool of candidate documents to be judged.
This pool consists of documents from all four categories that received a \emph{\texttt{view record}}, \emph{\texttt{download}} or \emph{\texttt{export}} user interaction after the query was issued, providing a strong implicit signal of potential relevance (as reported in~\cite{hienert2019digital}) and a tractable set for annotation.
% +Dhop
To reduce potential exposure bias and improve pool completeness, we further augmented the candidate set using BM25 and ColBERT with a supervised expanded form of the query, thereby incorporating retrieval-based candidates in addition to interaction-derived documents.
A detailed discussion of the pooling techniques employed is provided in the project GitHub repository.
% -Dhop

Relevance judgments were then automatically generated using an LLM on the selected interactions.
For our study, we employed OpenAI's \texttt{gpt-5-mini}, which was also used to generate descriptions and narratives for the topics. 
We developed a structured prompting strategy that provided the LLM with the topic description and the document metadata, instructing it to assess their relevance.
Judgments were made on a graded relevance scale from 0 to 4, defined as follows: `0' (Not Relevant), `1' (Marginally Relevant), `2' (Fairly Relevant), `3' (Highly Relevant), and `4' (Perfectly Relevant). 
% This fine-grained scale allows the use of more sophisticated evaluation metrics, such as Discounted Cumulative Gain~\cite{ndcg}.
Here, we opted against a Likert scale~\cite{likert}, as it measures subjective perception rather than objective topical alignment, and the $0-4$ scale is a TREC-adopted standard that integrates directly with IR evaluation metrics like nDCG~\cite{ndcg}.
Following this strategy, we finally obtain a pool of $85,158$ LLM-annotated relevance judgments from four distinct categories for $200$ topics. 
The average number of relevant documents per category is $114.4$, $121.94$, $48.44$, and $14.84$ for \texttt{publications}, \texttt{research data}, \texttt{variables}, and \texttt{instruments \& tools} , respectively.
%The average number of relevant documents per query in each category is shown in Table \ref{tab:reldocs}.
% In order to establish the trustworthiness of the LLM judgements, we verified a randomly chosen sample of 40\% of the total annotations by field experts and the outcome was in favour of LLM judgements.
%We established the trustworthiness of our LLM-based relevance judgments through expert validation. An evaluation of a randomly chosen 10\% sample of the annotations by field experts revealed total agreement, confirming the LLM's assessments.
%We tested the quality of the LLM-based relevance judgments through expert validation. Five randomly selected topics (5\% of all topics) with 10 documents in each of the 4 categories were independently evaluated by two authors using the 4-point relevance scale from above. Agreement between the human judgments and the LLM annotations was measured using quadratic-weighted Cohen’s $\kappa$, and we report the average agreement across authors. The resulting agreement ($\kappa = 0.X$) indicates substantial agreement.

To validate the LLM-generated judgments, four human experts independently annotated all pooled documents from twenty randomly selected topics (10\% of the total), using the same graded relevance scale (0–4).
% To validate the LLM-generated judgments, two human experts independently annotated documents from ten (10\% of total topic set) different randomly selected topics using the same graded relevance scale (0-4). 
Agreement between the human and the LLM judgments was measured using quadratic-weighted Cohen’s $\kappa$~\cite{Cohen1968}, yielding $\kappa = 0.86$, which indicates substantial agreement~\cite{manning2008introduction}.
In cases of disagreements, differences were predominantly within a single relevance level (i.e., $|score_{LLM}(Q, D) - score_{Human}(Q, D)| = 1$). These discrepancies typically occurred in borderline or subjective cases where either assessment was defensible, suggesting that the LLM’s judgments fall within an acceptable range of human interpretation.
% Furthermore, we observed that in cases of disagreement, the discrepancy was mostly within a single scale point, that is, $|score_{LLM}(Q, D) - score_{Human}(Q, D)| = 1$, where the individual scores respectively denote the relevance score assigned by the LLM and human expert. These differences occurred in subjective borderline cases where either assessment was justifiable, indicating that the LLM's judgments fell within acceptable bounds of human interpretation.

%Since our proposed \dataset supports multi-categorical search intent, unlike traditional TREC-formatted evaluation, one of the vital evaluation criteria here is the document category. 
Unlike traditional TREC-style evaluations, our proposed dataset is designed to support category-aware search intents, making the document category a key evaluation criterion.
Therefore, each judgment in the \dataset dataset is enriched with a `category' field, specifying the resource type of the judged document. This unique feature enables not only overall performance analysis but also a detailed examination of retrieval effectiveness by category, which is important for understanding an integrated system’s strengths and weaknesses across different types of resources. 

\section{Benchmark Analysis and Validation}

\subsection{Baselines} \label{subsec:baseline}

To establish a comprehensive performance baseline for the MIRA dataset, we evaluated a range of retrieval models spanning key IR paradigms. 
Our evaluation encompasses a range of techniques, from traditional term-based matching to advanced neural architectures. 
Specifically, we employed (i) the classic BM25~\cite{bm25,bm25_beyond} as a strong lexical baseline, (ii) the Relevance-based Language Model, RLM~\cite{rlm_sigir01,rlm} to represent traditional pseudo-relevance feedback and query expansion, (iii) the late-interaction neural model, ColBERT~\cite{colbert_sigir20} for efficient contextualized retrieval, and (iv) the SOTA model inspired by sequence-to-sequence transformer, MonoT5~\cite{monot5}. 
% This selection ensures our benchmark is tested against a diverse and representative set of approaches, providing a comprehensive foundation for future comparative research on multi-categorical retrieval. 
% + Added
% \textcolor{blue}{

We evaluate models in a category-aware setting, assessing effectiveness within each document category to analyze performance differences across heterogeneous resource types and identify category-specific strengths and weaknesses.
% }
We intentionally exclude LLM-based retrievers from our baseline comparison, as the significant inference latency and resource costs associated with these models \cite{ma-etal-2025-drama,DBLP:journals/corr/abs-2505-12260,li2026mixlmhighthroughputeffectivellm} preclude their deployment in real-time search scenarios, including the operational context of the GESIS Search platform.
The hyperparameter settings for baseline models, as well as topic modeling, are available in the GitHub repository to support reproducibility. 
% - Added

% The baseline BM25$_{desc}$ in Table~\ref{tab:result} is a variation of BM25 in which the query description (i.e., \texttt{<desc>} in Figure~\ref{fig:topic}) is used alongside the \texttt{<title>} during retrieval.
%The baseline, BM25$_{desc}$ in Table \ref{tab:result} is a variation of BM25 where the description of the query, i.e. `<des>' (see Figure~\ref{fig:topic}) information is leveraged, together with the `<title>', during retrieval.
% These models range from traditional term-matching to modern neural architectures, allowing us to benchmark performance across different technical approaches.
% Specifically, we employ the following models:
% \begin{itemize}
%     \item BM25~\cite{bm25,bm25_beyond}: 
%     \item {RLM}~\cite{rlm,rlm_sigir01}: 
%     \item {ColBERT}~\cite{colbert_sigir20}: 
%     \item {MonoT5}~\cite{monot5}: 
% \end{itemize}

% \paragraph{BM25}\cite{bm25,bm25_beyond}

% \paragraph{RLM}\cite{rlm,rlm_sigir01}

% \paragraph{ColBERT}\cite{colbert_sigir20}

% \paragraph{MonoT5}\cite{monot5}

\subsection{Evaluation}

We report four complementary metrics: (i) P$@10$ captures early precision, reflecting user expectations for first-page results in real-time systems, (ii) nDCG$@10$ accommodates our graded relevance scale by rewarding highly relevant documents, i.e. more than marginally relevant ones~\cite{ndcg}, (iii) MAP serves as our primary overall effectiveness measure across the full ranked list, and (iv) GMAP addresses query-level performance variability; unlike arithmetic mean, it is sensitive to low-performing queries and penalizes catastrophic failures on specific information needs~\cite{robertson2006gmap}.
Given the number of topics and multiple pairwise comparisons across categories and metrics, we apply a conservative significance threshold of $p<0.0001$ with paired t‑tests to reduce the likelihood of Type I errors.

% 
% 
% The models are evaluated using standard IR metrics, such as Precision at $5$ (P$@5$) and $10$ (P$@10$), Normalized Discounted Cumulative Gain at $5$ (nDCG$@5$) and $10$ (nDCG$@10$), and Mean Average Precision (MAP). 
The results are summarized in Table~\ref{tab:result}.
% We report the performance evaluation of the baseline models as mentioned in Section \ref{subsec:baseline} on all four categories (i.e. \pub, \rd, \var~and \ins), which are summarized in Table \ref{tab:result}.
The variation in performance across categories showcases the uniqueness of our integrated, category-sensitive search task, where each category provides a separate search space for the same user query. 
This means that for a particular user query, retrieval outcomes are likely to uncover multiple facets of the query depending on its search intent and the specific category, and hence our proposed benchmark provides a multi-faceted evaluation paradigm. 
The results in Table \ref{tab:result} clearly show that \pub~exhibit the highest retrieval effectiveness in all four evaluation metrics. 

% From Table \ref{tab:result}, a clear performance hierarchy is evident across the models. 
% from the results. 
% The classic bag-of-words model,  BM25, serves as a strong but lower-performing baseline.
% Extending the scope of the query by including the description to the short `title' further improves the performance.
% This highlights the value of richer textual representations even for bag-of-words-based models.
% The improvement (not always significant) is further evident with the pseudo-relevance feedback (RLM).
%  -- 
% The neural and le -- rning-based models significant
% ly outperform the statistical baselines.
% While the late interaction-based ColBERT exhibits superiority over other models - across all metrics both in the \pub~and \ins~categories, on the other hand, the sequence-to-sequence model, MonoT5 dominates in the \rd~category, and partially in \var.
% MonoT5, that frames ranking as a text generation task, achieves the highest scores across all three metrics, in the \rd~category with an nDCG$@10$ of $0.5736$, a MAP of $0.4347$, and $0.3297$ as GMAP.
% % Its superior performance demonstrates the model’s
% The superior performance of neural SOTA models demonstrate their ability to capture the global relevance relationship between queries and documents across diverse categories in MIRA dataset.

% \textcolor{red}{
Table~\ref{tab:result} presents a clear, statistically validated performance hierarchy that holds consistently across all categories and metrics. 
The classic bag-of-words based BM25 provides a strong lexical baseline, while 
% RLM delivers incremental, though not always significant, improvements via pseudo-relevance feedback. 
RLM shows significant improvements over BM25 
% (with paired t‑tests confirming significance at $p<0.0001$) 
in several cases (marked as $\dagger$~in Table~\ref{tab:result}), particularly for \texttt{Publications} and selected \texttt{Research Data} categories.
Crucially, ColBERT and MonoT5 both achieve significant advancements over RLM across nearly all metrics and categories (`*'-ed in Table~\ref{tab:result}), highlighting the efficacy of neural approaches. 
% Given the number of topics and multiple pairwise comparisons across categories and metrics, we apply a conservative significance threshold of  $p<0.0001$ with paired t‑tests to reduce the likelihood of Type I errors.
While ColBERT -- employing late interaction -- demonstrates slightly more consistent superiority, particularly across all metrics in \pub~ and \ins, MonoT5 leads in \rd~ and partially in \var, attaining peak scores in this category (nDCG$@10$: $0.5736$, MAP: $0.4337$, and GMAP: $0.3297$). 
These results confirm that state-of-the-art neural models effectively capture global query–document relevance across diverse collection types in the MIRA dataset.
% }

These results demonstrate two key points. 
\emph{First}, the MIRA dataset can discriminate between retrieval models of varying sophistication, revealing statistically significant performance gaps that align with expected trends in the field.
\emph{Second}, the task of multi-categorical retrieval in a scholarly domain remains challenging, with the highest MAP score below $0.6$.\footnote{Note that, this performance ceiling is notably higher than on traditional TREC ad-hoc collections, where MAP scores typically plateau between $0.25 - 0.40$~\cite{monot5,colbert_sigir20,monoduo}. 
We attribute this to the rich, structured metadata in scholarly objects and the presence of explicit categorical signals in real user queries. 
Crucially, however, substantial headroom remains for cross-category and category-aware retrieval challenges.}

MIRA, thus, offers a rich basis for developing and evaluating more advanced integrated and category-aware search systems.
% 
% \textcolor{blue}{
The observed performance differences across categories highlight the importance of category-aware modeling. Differences in document length, metadata richness, and vocabulary distribution suggest that ranking functions optimized for one category may not transfer directly to others. MIRA therefore provides a controlled setting for studying category-sensitive retrieval strategies.
% }

\begin{table*}[!t]
\centering
\caption{
\small
% Performance evaluation of both statistical retrieval models and neural re-rankers on our proposed \dataset dataset.
Performance comparison of both statistical and neural retrieval models on the MIRA dataset. 
Results are reported using P@10, nDCG@10, MAP, and GMAP. 
The best performance across the metric-category pair is highlighted in bold.
Both neural models are significantly better (paired t-test with $p < 0.0001$) than BM25.
Superscript $^\dagger$ indicates a significant difference between BM25 and RLM.
Significant improvements over RLM by the two neural models are 
% determined using a paired t-test ($p < 0.0001$) and are 
highlighted with `*'.
}
\begin{adjustbox}{width=0.9\textwidth}
\begin{tabular}
{l | @{~~}c@{~~}c@{~~}c@{~~}c | @{~~}c@{~~}c@{~~}c@{~~}c | @{~~}c@{~~}c@{~~}c@{~~}c | @{~~}c@{~~}c@{~~}c@{~~}c}

\toprule

& \multicolumn{4}{c}{\textbf{\pub}}
& \multicolumn{4}{c}{\textbf{\rd}}
& \multicolumn{4}{c}{\textbf{\var}} 
& \multicolumn{4}{c}{\textbf{\ins}} \\

\cmidrule{2-17}

% & P & nDCG & \multirow{2}{*}{MAP} & \multirow{2}{*}{GMAP} 
% & P & nDCG & \multirow{2}{*}{MAP} & \multirow{2}{*}{GMAP} 
% & P & nDCG & \multirow{2}{*}{MAP} & \multirow{2}{*}{GMAP} 
% & P & nDCG & \multirow{2}{*}{MAP} & \multirow{2}{*}{GMAP} \\
% % 
% & @10 & @10 & & 
% & @10 & @10 & &
% & @10 & @10 & &
% & @10 & @10 & & \\

& P@10 & nDCG & \multirow{1}{*}{MAP} & \multirow{1}{*}{GMAP} 
& P@10 & nDCG & \multirow{1}{*}{MAP} & \multirow{1}{*}{GMAP} 
& P@10 & nDCG & \multirow{1}{*}{MAP} & \multirow{1}{*}{GMAP} 
& P@10 & nDCG & \multirow{1}{*}{MAP} & \multirow{1}{*}{GMAP} \\
\cmidrule{2-17}

\textbf{BM25}
& 0.6120 & 0.6091 & 0.5098 & 0.5260
& 0.5045 & 0.5321 & 0.4023 & 0.3058
& 0.4905 & 0.4408 & 0.5057 & 0.1648
& 0.4190 & 0.4711 & 0.4540 & 0.2152 \\

\textbf{RLM}
& 0.6242$^\dagger$ & 0.6213$^\dagger$ & 0.5200$^\dagger$ & 0.5365$^\dagger$
& 0.5146 & 0.5427$^\dagger$ & 0.4103 & 0.3119$^\dagger$
& 0.5003 & 0.4496 & 0.5158$^\dagger$ & 0.1681 
& 0.4274 & 0.4805 & 0.4631 & 0.2195 \\

\textbf{ColBERT}
& \textbf{0.6523}$^*$ & \textbf{0.6492}$^*$ & \textbf{0.5434}$^*$ & \textbf{0.5607}$^*$
& \textbf{0.5377}$^*$ & 0.5639$^*$ & 0.4263$^*$ & 0.3241$^*$
& 0.5198$^*$ & 0.4672$^*$ & \textbf{0.5354}$^*$ & \textbf{0.1745}$^*$
& \textbf{0.4436}$^*$ & \textbf{0.4988}$^*$ & \textbf{0.4807}$^*$ & \textbf{0.2243}$^*$ \\

\textbf{MonoT5}
& 0.6365$^*$ & 0.6335 & 0.5302$^*$ & 0.5470$^*$
& 0.5247$^*$ & \textbf{0.5736}$^*$ & \textbf{0.4337}$^*$ & \textbf{0.3297}$^*$
& \textbf{0.5288}$^*$ & \textbf{0.4752} & 0.5249$^*$ & 0.1711$^*$
& 0.4349$^*$ & 0.4890 & 0.4713$^*$ & 0.2234$^*$ \\

% \textbf{RLM} 
% & 0.6334 & 0.6038 & 0.6170 & 0.6213 & 0.5200
% & 0.5495 & 0.5146 & 0.5674 & 0.5427 & 0.4103
% & 0.5375 & 0.5003 & 0.4483 & 0.4496 & 0.5158
% & 0.5090 & 0.4274 & 0.4533 & 0.4805 & 
% 0.4631
% \\

% \textbf{ColBERT}
% & \textbf{0.6619} & \textbf{0.6310} & \textbf{0.6448} & \textbf{0.6492} & \textbf{0.5434}
% & 0.5709 & 0.5347 & 0.5896 & 0.5639 & 0.4263
% & \textbf{0.5580} & \textbf{0.5193} & \textbf{0.4653} & \textbf{0.4667} & \textbf{0.5354}
% & \textbf{0.5202} & \textbf{0.4368} & \textbf{0.4633} & \textbf{0.4911} & \textbf{0.4733}
% \\

% \textbf{MonoT5}
% & 0.6458 & 0.6157 & 0.6291 & 0.6335 & 0.5302
% & \textbf{0.5807} & \textbf{0.5439} & \textbf{0.5997} & \textbf{0.5736} & \textbf{0.4337}
% & 0.5470 & 0.5091 & 0.4562 & 0.4576 & 0.5249
% & 0.5180 & 0.4349 & 0.4613 & 0.4890 & 0.4713
% \\

% BM25$_{des}$ & 0.3576 & 0.2802 & 0.3472 
% & 0.4080 & 0.3296 & 0.4387 
% & 0.3570 & 0.2997 & 0.3792
% & 0.3568 & 0.2707 & 0.3453 \\

% RLM & 0.3678 & 0.2883 & 0.3572 
% & 0.4197 & 0.3390 & 0.4512 
% & 0.3910 & 0.3082 & 0.3901 
% & 0.3670 & 0.2784 & 0.3552 \\

% ColBERT & 0.3917 & 0.3069 & 0.3803
% & 0.4469 & 0.3611 & 0.4805 
% & 0.3917 & 0.3282 & 0.4153
% & 0.3908 & 0.2965 & 0.3782 \\

% MonoT5 & 0.4051 & 0.3196 & 0.4036
% & 0.4586 & 0.3704 & 0.4930 
% & 0.4012 & 0.3368 & 0.4261
% & 0.4011 & 0.3096 & 0.3991 \\

\bottomrule

\end{tabular}
\label{tab:result}

\end{adjustbox}
\end{table*}

\section{Discussion}

% We begin this section by highlighting one limitation 
% We now consider the implications and context of our work in constructing the MIRA benchmark. 
% First, we address a key methodological limitation regarding implicit relevance signals, which informed our assessment choice. 
% Next, we contextualize our contribution by formally distinguishing the proposed dataset from classic TREC ad-hoc datasets. 
% Finally, we conclude by exploring the potential applications and research tasks enabled by this new dataset.

We now contextualize the MIRA benchmark by addressing three key aspects: (i) a methodological limitation regarding implicit relevance signals and their impact on our assessment approach, (ii) a formal distinction between our dataset and classic TREC ad-hoc collections, and (iii) the novel applications and research tasks enabled by this resource.

% \subsection{Limitations of Implicit Relevance Signals}

\paragraph{\textbf{Limitations of Implicit Relevance Signals.}}
As reported in~\cite{hienert2019digital}, user interactions, such as \emph{\texttt{view record}}, \emph{\texttt{download}}, or \emph{\texttt{export}}, are valuable implicit signals for inferring relevance. However, it is important to note that these signals may not be definitive. Our analysis confirms that these interactions do not always correlate perfectly with topical relevance. For instance, a user may view a record to quickly ascertain its irrelevance, or download an item for administrative purposes rather than for its topical content. This potential for noise and misinterpretation motivated our decision to rely on explicit, LLM-assisted human assessments to establish the ground-truth relevance judgments in the MIRA benchmark, thereby ensuring greater accuracy.

% \subsection{Disti\textbf{nction from Classic TREC Ad-Hoc Datasets.}}

% While \dataset is built upon the foundational Cranfield paradigm shared by classic TREC ad-hoc datasets, it represents a significant evolution to address the complexities of modern, integrated search.

\paragraph{\textbf{Distinction from Classic TREC Ad-Hoc Datasets.}}
The proposed \dataset dataset represents a significant evolution from classic TREC ad-hoc test collections, which were largely designed for evaluating retrieval over a uni-categorical document corpora.
In contrast, \dataset items are fundamentally multi-categorical, integrating four distinct scholarly resource types~--~\pub, \rd, \var, and \ins~--~within a single benchmark, thereby reflecting the reality of modern, integrated search platforms.
Unlike traditional benchmark collections (e.g., TREC) that rely mostly on manually curated evaluation topics, MIRA derives its topics from real user interactions.
In addition, while TREC depends on fully human-generated relevance judgments, \dataset employs a scalable LLM-assisted framework to produce graded relevance assessments across heterogeneous resource categories.
% Furthermore, while TREC relied on cosmetic topics and expensive manual relevance judgments, \dataset is grounded in real user queries and uses a scalable, LLM-assisted methodology to generate graded relevance assessments across all categories, addressing the critical gap in evaluating complex, cross-categorical information retrieval.
The key differences are summarized in Table~\ref{tab:trec_comparison}.
%The differences are multifaceted and fundamental, as summarized in Table~\ref{tab:trec_comparison}.

\begin{table}[!t]
\centering
\caption{
\small
Comparisons between Classic TREC Ad-hoc datasets and the MIRA dataset.
}
\begin{adjustbox}{width=0.9\columnwidth}
\begin{tabular}
{p{0.23\columnwidth}|p{0.28\columnwidth}|p{0.49\columnwidth}}
\hline
\textbf{\centering Features}    & \textbf{\centering TREC Ad-Hoc}       & \textbf{\centering \dataset}                                      \\ \hline
\textbf{Document \newline Homogeneity} &
  Homogeneous \newline (e.g., newswire articles, web pages). &
  Intentionally heterogeneous (\rd, \var, \pub, \ins). \\ \hline
\textbf{Relevance \newline Criteria} &
  Uniform across the collection (textual aboutness). &
  Category-dependent \newline (e.g., reusability for data, measurability for variables, methodological guidance for tools). \\ \hline
\textbf{Topic Origin}         & Expert-created, simulated.                    & Real user queries from a live search platform.                    \\ \hline
\textbf{Judgment Scope}       & Focused on a single, primary document type.   & Encompasses multiple, distinct document types for a single topic. \\ \hline
% \textbf{Judgment Methodology} & Exclusively  human assessors.                  & LLM-assisted, scaled via automated prompting.                     \\ \hline
\textbf{Core Evaluation Goal} & Ranking effectiveness within a single corpus. & Integrated ranking effectiveness across multiple corpora.         \\ \hline
\end{tabular}
\label{tab:trec_comparison}
\end{adjustbox}
\end{table}

% \subsection{Application of the MIRA dataset}
\paragraph{\textbf{Application of the MIRA dataset.}}
This new resource opens up numerous avenues for future research.
Primarily, it serves as a benchmark for evaluating integrated, category-aware, and cross-category retrieval systems, aiming to produce a single, effective ranking across heterogeneous item types.
It can also be used to study category-specific retrieval models, investigating, for instance, whether a specialized model for {\var} outperforms a generic one.
Furthermore, it enables researchers to explore additional tasks, such as \emph{result diversification} (ensuring that top results span multiple relevant categories), \emph{fairness in ranking} (analyzing whether certain categories are systematically under-represented), and \emph{query performance prediction} across different corpora.
Finally, its foundation on a dual-lingual platform invites research on {code-switching and cross-lingual retrieval}.

% \inote{Graded relevance?}
% \inote{Other than traditional retrieval, mention what are some possible research that can use this dataset.}

% The \dataset dataset also serves as a foundation for integrating offline and online evaluation approaches. The continuous evaluation approach in IR \cite{keller2025} combines methods for assessing and improving search systems in dynamic environments. For example, systems like GESIS Search are continuously evolving in terms of new users, new topics and queries, new documents in different categories, and new search system functionalities and improvements. Due to its automated creation workflow, \dataset serves as a valuable evaluation resource that can be continuously expanded with new queries, topics, documents, and user interactions for relevance and beyond. This creates a living evaluation corpus that remains continuously aligned with the evolving system and its user base. Since we have demonstrated an automated creation process for this dataset using user logs alongside automatic extraction and enrichment methods, continuous evaluation datasets can now be generated for any domain and search system.

The \dataset dataset enables the integration of offline and online evaluation in IR. Continuous evaluation~\cite{keller2025} assesses and improves search systems in dynamic environments where users, topics, and documents evolve. Platforms like GESIS Search exemplify this need. With its automated creation workflow, MIRA can be continuously expanded with new topics, documents, and user interactions, forming a living evaluation corpus that stays aligned with system and user evolution. By automating dataset generation from user logs and metadata enrichment, our approach enables continuous evaluation datasets to be produced for any domain or search system.

\section{Conclusion and Future Work}

The modern search experience is integrated, yet IR benchmarks have lagged behind, constrained by a lack of collections that mirror this reality.
We introduce MIRA, a novel benchmark designed to address the critical evaluation gap in multi-categorical information retrieval.
The \dataset dataset directly confronts this challenge by providing a unified framework encompassing four distinct scholarly categories -- \pub, \rd, \var, and \ins\;-- all grounded in real user queries from the GESIS Search platform.
Our primary contribution is a resource that is both realistic in origin and novel in its construction, using LLMs to efficiently produce topic narratives and scalable, graded relevance judgments spanning all categories.

%comment: took out this section for discussion 7.2
%make a short point to this section here later. 1-3 sentences.
The \dataset dataset provides a comprehensive benchmark for evaluating integrated, category-aware and cross-category retrieval systems,
% \textcolor{blue}
% {
% We introduce MIRA, a benchmark designed to evaluate category-aware ranking in an integrated scholarly search environment. Unlike traditional monolithic collections, MIRA 
spanning over multiple heterogeneous resource types.
It supports research on category-aware ranking models and enables analysis of performance differences across heterogeneous resource types, along with investigations into result diversification, fairness-aware ranking, and query performance prediction.
% }
Notably, its automatic generation process lays the groundwork for a continuous evaluation paradigm, in which evolving systems and their user base are captured within a living evaluation corpus.

Looking ahead, we plan to expand the \dataset dataset by increasing the number of judged topics and increasing the depth of the pool of relevance assessments, potentially through continued LLM-assisted methods complemented by human-in-the-loop validation.
We also envision organizing a shared task or benchmark challenge to foster community engagement and establish baseline performance metrics.
We believe \dataset provides a foundational step towards evaluating the next generation of search systems and will inspire further work in complex, multi-faceted, and federated information retrieval.

%Extending the topic set, we plan to enrich the benchmark to support the task of scholarly graph completion in a future release, where systems would leverage the interlinked metadata to predict missing relationships, such as automatically suggesting relevant research data for a given publication or identifying which variables are measured by a specific survey instrument.

% For a future release, we plan to enrich the benchmark to support the task of scholarly graph completion, where systems would leverage the interlinked metadata to predict missing relationships, such as automatically suggesting relevant research data for a given publication or identifying which variables are measured by a specific survey instrument.

%%% Local Variables:
%%% mode: latex
%%% TeX-master: main
%%% End:

\begin{acks}
This research has been funded by the Deutsche Forschungsgemeinschaft (DFG, German Research Foundation), project number 407518790, and partly funded by Horizon Europe grant OMINO (grant number 101086321).
Also, this publication received partial support from the European Research Council (ERC) under the Horizon 2020 research and innovation program (Grant agreement No. 884951), and Research Ireland to the Insight Centre for Data
Analytics under grant no. 12/RC/2289\_P2. 
\end{acks}

\bibliographystyle{ACM-Reference-Format}
\balance
\bibliography{references}

\end{document}